\documentstyle[preprint,aps]{revtex} 
\tightenlines
\begin{document}
\draft
\renewcommand{\thesection}{\arabic{section}} 
\renewcommand{\theequation}{\arabic{equation}} 

\newcommand{\beq}{\begin{equation}}
\newcommand{\eeq}{\end{equation}}
\newcommand{\bea}{\begin{eqnarray}}
\newcommand{\eea}{\end{eqnarray}}

\preprint{IFIC/97--107}

\title{
State-dependent Jastrow correlation functions for $^4$He nuclei 
}

\author{
R.~Guardiola 
}
\address{
Dept.\ de F\'{\i}sica At\'omica, Molecular i Nuclear, 
Universitat de Val\`encia, \\ 
Avda.~Dr.~Moliner 50, E-46100 Burjassot (Valencia), Spain
}
\author{
M.~Portesi\thanks{
E-mail address: mariela.portesi@uv.es} 
}
\address{
IFIC (Centre mixt: CSIC -- Universitat de Val\`encia), \\
Avda.~Dr.~Moliner 50, E-46100 Burjassot (Valencia), Spain 
} 

\maketitle

\begin{abstract}

We calculate the ground-state energy for the nucleus $^4$He with V4
nucleon interactions, making use of a Jastrow description of the
corresponding wavefunction with {\em state-dependent} correlation factors.
The effect related to the state dependence of the correlation is quite
important, lowering the upper bound for the ground-state energy by some
2~MeV.

\end{abstract}

\vspace{2cm}

The physical problem of describing many interacting identical particles
from a microscopic point of view can be attacked using a number of
techniques. A frequent {\sl ansatz} for the ground state (g.s.) of a
many-body system is the Jastrow wavefunction. For a nuclear system the
Jastrow method describes the wavefunction in terms of the product of
two-body correlations between all pairs of nucleons, acting upon a
suitable reference state. This approach has been very fruitful in the
description of extended strongly interacting systems, such as helium
liquids at zero temperature and even nuclear and neutron matter.

The practice of the Jastrow correlated method in infinite systems requires
the use of massive resummation techniques, such as the hypernetted-chain
(HNC) algorithms~\cite{FR_NC25a,K_NPa317} or Monte Carlo (MC) sampling
procedures~\cite{McM_PRa138}. After more than 20 years of practice this
method has proven to be perhaps the best variational description of
many-particle systems.

A different question is the treatment of finite strongly interacting
systems, the main difficulty being related to the $A$-body interparticle
correlation induced by the localization of the system. From the point of
view of massive resummation of diagrams \`a la HNC, the most unpleasant
fact is that any diagram becomes {\em elementary}, which in turn makes the
HNC resummation very involved. Even though the formal equations were
written many years ago~\cite{FR_NPa328}, it is only quite recently that
some results for finite nuclei with model, realistic-like, interactions
have been presented~\cite{fabro}. On the other hand, there is no special
difficulty, not present in the case of extended systems, with regard to
the use of variational MC methods. In any case, the presence of
state-dependent interactions and, in turn, of state-dependent correlations
gives rise to a formidable computational complexity.

An alternative way to calculate expectation values for Jastrow
wavefunctions consists of the consideration of {\em cluster expansions},
in which the mean value of an operator for an $A$-particle system is
obtained by means of nonlinear expansions of the expectation value for
$n$-body subsystems. As in the case of HNC and MC methods, the basic idea
is taken from statistical mechanics. Cluster expansions adapted to finite
systems were analysed by Clark and Westhaus~\cite{CW_JMP9} and used in
nuclei by one of us~\cite{G_NPa328}. MC studies~\cite{BBG_PLb198}, as well
as the recent HNC analysis~\cite{fabro}, have confirmed the quality of
these expansions.

Previous calculations of the binding energy for several light nuclei make
use of state-independent~(SI) Jastrow correlations between the nucleons.
Our aim is to study the effect of dealing with {\em state-dependent}~(SD)
correlation functions within a Jastrow variational context. As an initial
step we will concentrate in the simple case of $^4$He, as well as in
simple central (but spin--isospin dependent) forces of semirealistic
nature. Such interactions are usually termed as V4, the number 4 referring
to the four two-body operators $\{\hat{\openone},\,
\vec\sigma_i\cdot\vec\sigma_j,\, \vec\tau_i\cdot\vec\tau_j,\,
(\vec\sigma_i\cdot\vec\sigma_j) (\vec\tau_i\cdot\vec\tau_j)\}$. The main
difference with respect to variational MC studies is that we will carry
out all calculations by means of algebraic techniques, avoiding the
presence of statistical errors. This will be very important when dealing
with heavier nuclei.

\hfill

The trial wavefunction, formally written as $|\Psi\rangle=\hat F {\cal A}
|\Phi\rangle$, is the product of a Jastrow correlation factor $\hat F$ and
an uncorrelated shell-model state. The latter is constructed with
single-particle orbitals from the harmonic-oscillator~(HO) potential, in
order to be able to exactly remove the spurious centre-of-mass~(CM)
motion. Denoting by $\{\phi_k(x_i),\,k=1,\ldots,A\}$ the set of relevant
normalized single-particle states (specified for the coordinates
$x_i\equiv(\vec r_i,\vec\sigma_i,\vec\tau_i)$ of the $i$th nucleon), the
antisymmetric reference state is nothing but the Slater determinant ${\cal
A}|\Phi\rangle = {\rm det}(\phi_k(x_i))/\sqrt{\!A!}$. The Jastrow factor
is the symmetrized~\cite{PW_NPa266,LP_NPa342} product of two-body
correlation functions, also of V4 structure, for all pairs of nucleons.
The number of factors that build up the total $\hat F$ is therefore
$N_C\equiv A(A-1)/2$. The correlation corresponding to a given pair $(ij)$
is expressed as
\beq
\hat f(x_i,x_j) \equiv \hat f_{ij} = 
f^c (r_{ij}) \, \hat{\openone} +
f^\sigma (r_{ij}) \, \hat P_\sigma(ij) +
f^\tau (r_{ij}) \, \hat P_\tau(ij) +
f^{\sigma\tau} (r_{ij}) \, \hat P_\sigma(ij) \hat P_\tau(ij)
\label{fij_gen} \eeq
where the spin and isospin dependence is explicitly written in terms of
spin or isospin exchange operators, and $f^c, f^\sigma, f^\tau$ and
$f^{\sigma\tau}$ are scalar functions of the interparticle distance
$r_{ij}\equiv |\vec r_i-\vec r_j|$. The SD correlated g.s.\ wavefunction
\`a la Jastrow is finally written, up to a missing normalization factor,
as
\beq
\Psi(x_1,\ldots,x_A)={\cal S}\left\{ \prod_{j>i=1}^A \hat f(x_i,x_j) \right\} 
{\cal A} \Phi(x_1,\ldots,x_A) 
\label{Psi} \eeq
where ${\cal S}$ is the symmetrizer for the operator inside the curly
brackets.

It is worth stressing that the antisymmetrization of $|\Phi\rangle$
---which is required to fulfil the fermionic nature of the nucleons--- as
well as the symmetrization in $\hat F$ ---imposed by the (in general)
non-commutative character of the different pair-correlation operators---
are sources of notable technical complications. From the symmetrization of
$\hat f_{12} \hat f_{13} \ldots \hat f_{A-1 A}$, we see that $N_C!$ terms
indeed constitute the correct, complete correlation. Then, as long as the
number of particles is small the problem remains manageable but its
difficulty rapidly grows with increasing $A$. In passing, let us mention
that in most works posing this sort of
treatment~\cite{LP_NPa342,O_AP118,CK_PRc32}, the so-called
independent-pair approximation\footnote{The independent-pair approximation
involves rewriting each $\hat f_{ij}$ as a scalar function of $r_{ij}$,
times an SD operator of the form $\hat{\openone}+\hat U_{ij}$, and then
retaining from the expansion of $\prod_{i<j} (\hat{\openone}+\hat U_{ij})$
only those terms with labels that correspond to independent pairs of
nucleons.} has widely been applied and so the question of symmetrization
has been avoided.

In order to obtain the g.s.\ energy we should compute the mean value of
the Hamiltonian
\beq
\hat H = -\frac{\hbar^2}{2m} \sum_{i=1}^A \nabla^2_i + 
\sum_{j>i=1}^A \hat V(x_i,x_j)
\label{ham} \eeq
where $\hbar^2/m= 41.47$~MeV~fm$^2$, and the potential $\hat V$ is a
two-body operator that depends on the relative spatial coordinate and the
spin--isospin degrees of freedom of each pair of nucleons. The
interactions employed here have V4 structure, which will be transformed to
an exchange-operator basis as in the case of the
correlation~(\ref{fij_gen}),
\beq
\hat V(x_i,x_j) = 
\sum_u g_u(r_{ij}) \sum_{p=1}^4 v_{u,p} \, \hat O_p(ij) 
\eeq
where $\hat O_1(ij)\equiv\hat{\openone}$, $\hat O_2(ij)\equiv\hat
P_\sigma(ij)$, $\hat O_3(ij)\equiv\hat P_\tau(ij)$ and $\hat
O_4(ij)\equiv\hat P_\sigma(ij) \hat P_\tau(ij)$. This representation is
very economical because in matrix form any exchange corresponds to a
sparse matrix with a single non-null element per row and
column~\cite{GMNBPW}. For our calculations we will consider the
quasirealistic potentials proposed by Afnan and Tang~\cite{AT,GFMP_NPa371}
and by Malfliet and Tjon~\cite{MT}, whose various components have Gauss
and Yukawa spatial dependence respectively, with strengths $v_{u,p}$ and
corresponding form factors $g_u^G(r)=\exp(-\mu_u r^2)$ and
$g_u^Y(r)=\exp(-\mu_u r)/r$. The Coulomb interaction will not be taken
into account in any case. Notice that the Hamiltonian~(\ref{ham}) is
symmetric under particle permutation and so is the operator $\hat
F^\dagger\hat H\hat F$. This property enables us to find the total kinetic
energy as $A$ times the corresponding value for a single particle, and the
total potential energy as $N_C$ times the contribution of a single pair of
nucleons.

\hfill 

The $^4$He results presented here can be obtained exactly by a direct
evaluation and further minimization of the mean value of the
Hamiltonian~(\ref{ham}) with a trial state of the form~(\ref{Psi}).
Nevertheless, we have followed a cluster representation with the purpose
of gaining some insight into this approach for future application to more
complicated situations, namely for heavier nuclei or more realistic
interactions. In that sense, the present example serves as a test for the
convergence of the cluster expansion when SD correlations are taken into
account. Besides, the approximation procedure is useful for seeking the
region where the minimizing parameters of the correlation functions should
be found. The methodology we followed to compute the required expectation
values is the one developed in~\cite{G_NPa328}, that we summarize below.

Consider an $n$-particle subsystem ($1\leq n\leq A$) whose Hamiltonian
operator $\hat H_n$ has the form~(\ref{ham}) with the replacement
$A\rightarrow n$. The $n$-body integrals are introduced as
\bea
J_n(\lambda) 
& = & \frac 1{A(A-1)\ldots(A-n+1)} \sum_{k_1,\ldots,k_n=1}^A 
      \langle \phi_{k_1}(x_1)\ldots\phi_{k_n}(x_n) | \nonumber \\ 
&   & {\cal S}\left\{ \prod_{i'<j'}^n \hat f_{i'j'} \right\} 
      \, e^{\lambda\hat H_n} \, 
      {\cal S}\left\{ \prod_{i<j}^n \hat f_{ij} \right\} \; 
      \sum_{{\rm P}=1}^{n!} \epsilon_{\rm P} \, 
| \phi_{k_1}(x_{{\rm P}1}) \ldots \phi_{k_n}(x_{{\rm P}n}) \rangle 
\eea
where the first sums extend over all the single-particle states relevant
for the given nucleus, and the last one runs over all permutations of
particle indices ($\epsilon_{\rm P}$ are their associated parities). The
matrix elements involve, besides spatial integrations over the whole
space, summations for the spin and isospin variables of the $n$ nucleons.
If one deals with spin/isospin saturated nuclei, these summations
translate into corresponding traces of products of exchange operators.
Their evaluation, although quite long, can be made once at the very
beginning of the calculations and the resulting values can be stored for
further use (a detailed explanation on this subject can be found
in~\cite{GMNBPW}). In the above equation, symmetrization of an $n$-body
correlation is required and therefore $n_C!\equiv (n(n-1)/2)!$ reorderings
are obtained (both on the left and on the right), in contrast with the
$N_C!$ terms arising from the complete $A$-particle problem. The quantity
we are interested in, the minimum value acquired by $\langle\Psi|\hat
H|\Psi\rangle/\langle\Psi|\Psi\rangle-E_{\rm CM}$, can be derived from the
$A$-body integral since
\beq
\langle\hat H\rangle = \left. \frac {d\ }{d\lambda} \ln J_A(\lambda) 
\right|_{\lambda=0} = \frac{J_A'(0)}{J_A(0)} . 
\label{ham_JA} \eeq
The second term in the g.s.\ energy takes proper account of the CM
movement and is given by the kinetic energy of a particle in the lowest HO
state, i.e.\ $E_{\rm CM} = 3\,\hbar^2\alpha^2/(4m)$, $\alpha$ being the HO
constant.

The binding energy should be obtained by scanning the space of trial
correlation functions. Except for light nuclei, this is impracticable with
present-day computers because of the huge number of evaluations to be
performed. It proves useful to introduce a multiplicative approximation
procedure by means of the {\em cluster integrals} ${\cal J}_\nu$ which are
defined by the recursion
\beq
J_n \equiv \prod_{\nu=1}^n {\cal J}_\nu^{n\choose \nu}
\label{Jnu} \eeq
for all $n=1,\ldots,A$. The interpretation of this chain of identities is
quite straightforward: one can obtain the $n$-body integral as a product
of different contributions arising from fewer-body interacting subsystems.
This idea allows us to decompose the energy expectation value as
\beq
\langle\hat H\rangle = 
E^{(1)} + \sum_{\nu=2}^A {A\choose\nu} \, \delta E^{(\nu)}
= A \, \frac{J_1'(0)}{J_1(0)} + \sum_{\nu=2}^A {A\choose\nu} \,
\left. \frac {d\ }{d\lambda} \ln {\cal J}_\nu \right|_{\lambda=0} . 
\label{ham_Enu} \eeq
For a nuclear system, one may assume that a reduced number, say $n$, of
cluster integrals would suffice in determining $J_A$ with great accuracy,
the remaining cluster integrals being close to unity so as to produce
negligible contributions $\delta E^{(\nu>n)}$ to the
energy~(\ref{ham_Enu}). The $A$-body integral calculated up to the $n$th
order turns out to be
\beq
J_A 
\simeq {\cal J}_1^A \ldots {\cal J}_n^{A\choose n} \; \equiv \; J_A^{(n)}
\eeq
with ${\cal J}_1,\ldots,{\cal J}_n$ deduced from the recursive
formulae~(\ref{Jnu}). Starting with $n=1$, one can systematically
construct a series of approximations for the energy by taking the
logarithmic derivatives of $J_A^{(n)}$. A detailed analysis of the
characteristics and convergence of this multiplicative cluster expansion
employing SI correlations has been carried out in previous
works~\cite{G_NPa328,BBG_PLb198}. Up to first order the energy has purely
kinetic origin and amounts to
\beq
E^{(1)} = 
\frac{\hbar^2 \alpha^2}m \left[ 3+5\left( \frac A4 -1 \right) \right] . 
\eeq
This expression is valid for $s$- and $p$-shell nuclei, namely $^4$He,
$^8$Be, $^{12}$C and $^{16}$O, for which the spatial single-particle
states are $\phi_{s}(\vec r)= (\alpha/\sqrt\pi)^{3/2}\exp(-\alpha^2
r^2/2)$ and $\{\phi_x,\phi_y,\phi_z\}_{p}(\vec r)= \sqrt 2\,\alpha\,
\{x,y,z\}\phi_{s}(\vec r)$. The first non-trivial result is obtained with
one- and two-body clusters:
\beq
E^{(2)} = E^{(1)} + {A\choose 2} 
\left[ \frac{J_2'}{J_2}-2\frac{{\cal J}_1'}{{\cal J}_1} \right]_{\lambda=0} . 
\eeq
Higher orders are constructed following similar steps, i.e.\ calculating
the successive corrections as suggested in equation~(\ref{ham_Enu}) so
that
\beq
E^{(n)} 
= E^{(n-1)}+ { A \choose n } \, \delta E^{(n)} . 
\eeq
Let us mention that at each order one can add at once the kinetic
contribution of only one of the $n$ particles and the potential due to a
given pair of them, with corresponding factors $n$ and $n_C$. The
approximation procedure will be performed up to a maximum of a few
clusters, studying the convergence of the partial results. Note that very
precise calculations are required, for two main reasons: the cancellations
in $\delta E^{(n)}$ implied by the definition of ${\cal J}_n$ (as the
$n$-body integral divided by all the ${\cal J}_{\nu <n}$), and the fact
that the statistical factors (explicitly factored out in the preceding
equation) introduce an enormous scaling with $n$.

\hfill

Let us now specify the form of the pair correlation functions. A very
useful, and rather easy to handle, parametrization is an expansion in a
set of Gaussians. Our {\sl ansatz} is the SD linear combination
\beq
\hat f_{ij} = \hat{\openone} + \sum_{m=1}^{N_\beta} 
e^{-\beta_m r_{ij}^2} \sum_{p=1}^4 a_{m,p}\,\hat O_p(ij) . 
\label{fij} \eeq
The correlation lengths $\beta_m$ and correlation depths $a_{m,p}$,
together with the HO constant $\alpha$, constitute a set of (real) free
parameters to be determined from energy minimization (throughout this
work, distances are measured in fm and the variables $\beta_m$ and
$\alpha$ will be given in fm$^{-2}$ and fm$^{-1}$ respectively). As it is
clearly seen, the SI problem arises as a particular case by setting
$a_{m,p\neq 1}=0$ and freely determining the $2 N_\beta+1$ variables
$\beta_m$, $a_{m,1}$ and $\alpha$.

In the case of $^4$He nuclei, the problem substantially simplifies due to
the fact that the reference state exhibits spatial symmetry. Indeed, the
uncorrelated g.s.\ is made up of the four particles in the same spatial
single-particle state, $\phi_{s}$. As is known, the operator $\hat P_{\vec
r}(ij) \hat P_\sigma(ij) \hat P_\tau(ij)$ characterizes the exchange of
all coordinates of the nucleons $i$ and $j$. Since $\hat P_{\vec r}(ij)$
leaves the helium wavefunction unaltered, the spin and isospin exchange
operators acting on ${\cal A}\Phi(x_1,x_2,x_3,x_4)$ give a change of sign,
and consequently the action of $\hat P_\sigma$ is equivalent to minus that
of $\hat P_\tau$. The same holds in second and third order of the cluster
expansion for $^4$He. Then, for this nucleus we are able to rewrite the
correlation factors with a simpler state dependence, the {\sl
ansatz}~(\ref{fij}) becoming
\beq
\hat f_{ij} = \hat{\openone} +
\sum_{m=1}^{N_\beta} e^{-\beta_m r_{ij}^2} 
[a_{m,c} \, \hat{\openone} + a_{m,\sigma} \hat P_\sigma(ij)]
\eeq
with central scalar and spin-exchange constituents only. The
state-independent study is performed, as usual, by setting
$a_{m,\sigma}=0$ for all $m$.

\hfill

In tables~\ref{tabS3} and~\ref{tabMT} we present the results corresponding
to the g.s.\ energy of the $^4$He nucleus, obtained using different trial
wavefunctions and nucleon--nucleon interactions. Table~\ref{tabS3}
corresponds to the Afnan--Tang S3 potential~\cite{AT} and its modified
version MS3~\cite{GFMP_NPa371}, while in table~\ref{tabMT} the
Malfliet--Tjon interactions MT~I/III and MT~V~\cite{MT} are considered.
The four-particle system has been characterized by a Jastrow prescription
with state-dependent correlations containing one and two Gaussian
components. For the sake of comparison, we also show in the tables the
concomitant state-independent energies (the SI values have earlier been
computed in~\cite{BBFG_JPg18} assuming $\hbar^2/m=41.50$~MeV~fm$^2$).
Together with the minima obtained in each case, we exhibit the optimum
values for the HO constant $\alpha$, correlation lengths $\beta_m$ and
correlation depths $a_{m,c}$ and $a_{m,\sigma}$, for $m\leq N_\beta$ with
$N_\beta=1$ and 2.

The S3 potential is parametrized in terms of five ranges $\mu_u$ and has
only singlet--even and triplet--even channels; the MS3 interaction
incorporates the odd ones. Although for helium both potentials yield the
same results employing SI pair correlation factors, the introduction of a
dependence upon the spin of the nucleons distinguishes between those
interactions. The MS3 shift with respect to the SI binding energy turns
out to be somewhat smaller in magnitude than the S3 one: $3.7\%$ ($9.1\%$)
against $3.9\%$ ($10.0\%$) respectively, for one (two) correlation
lengths. This difference is related to the saturation properties of the
modified Afnan--Tang interaction.

The potentials proposed by Malfliet and Tjon are a superposition of an
attractive and a repulsive Yukawa component. In the MT~I/III case, the
spin-singlet and spin-triplet even channels are split; for the MT~V case
it is assumed that these two forces can be replaced by an average
effective potential which is identical in both channels. The last
situation allows one to treat the four nucleons as identical spinless
bosons, meaning that the SI and SD approaches for the MT~V potential give
the same result, which is also consistent with the state-independent
treatment for the MT~I/III force. We find that with this interaction the
state-dependent binding energy diminishes by almost $1\%$ ($3.7\%$) of the
SI value for $N_\beta=1$ ($N_\beta=2$).

The results presented here can be compared with other existing
calculations, among which we may mention the energies given using MC
techniques and the spin--isospin-dependent translationally invariant
coupled cluster (TICI2) treatment~\cite{GMNBPW}. In the case of the MT~V
potential, for instance, our ($N_\beta=2$)-result is very close to the
Green function MC value of $(-31.3\pm 0.2)$~MeV \cite{ZK_NPa356} and the
diffusion MC energy, $(-31.32\pm 0.02)$~MeV \cite{BBFG_JPg18}. After
performing a coupled cluster calculation for the alpha particle,
Zabolitzky~\cite{Z_PLb100} arrived at the conclusion that the g.s.\
eigenvalue of the MT~I/III Hamiltonian should be $(-33.4\pm 0.1)$~MeV. The
TICI2 method with SD linear correlations of the V4 type yields the
following results~\cite{GMNBPW} for the quoted potentials: $-28.19$~MeV
(S3), $-27.99$~MeV (MS3), $-30.81$~MeV (MT~I/III) and $-29.45$~MeV (MT~V).
These values have not been optimized with respect to the HO parameter and
hence they may be slightly underestimated in magnitude. It can be seen
that the results given here are comparable with the binding energies of
the TICI2 method; moreover, our results with two betas are lower than the
latter by more than 1~MeV.

In order to look at the convergence of the above mentioned multiplicative
cluster expansion, we show the partial results corresponding to the
minimum energies found with one and two $\beta$ parameters.
Table~\ref{tabdelt1} contains for $N_\beta=1$ the first-order energy
$E^{(1)}$ as well as the concomitant corrections up to the second, third
and fourth orders, which should be multiplied by the statistical factors
6, 4 and 1 respectively. One can see that in this situation the SI
fourth-order corrections are much smaller than the respective third-order
terms, and using SD correlation functions $|\delta E^{(4)}|$ remains lower
than $|\delta E^{(3)}|$; finally, the third- and fourth-order corrections
have opposite signs. The influence of the introduction of a second SD
correlation component is shown in table~\ref{tabdelt2}, giving evidence
that in this case four-body SD contributions gain in importance and thus
are not negligible. Moreover, for the MT potentials the third- and
fourth-order corrections both contribute to lower the second-order energy.
We stress that, at this step, our purpose is to compare the minimum total
energy with the third-order approximation obtained assuming the optimum
parametrization of tables~\ref{tabS3} and~\ref{tabMT}, which in principle
should not be equal to the minimum $E^{(3)}$.

In conclusion, let us remark on the good agreement of our results with the
available MC ones, when we take into account just two correlation
components. We have shown the importance of introducing a dependence with
the spin and isospin of the nucleons in the pair correlation functions,
the shift in the energies with respect to the state-independent values
being of around 1--10\%, depending on the two-body interaction considered.
As a side benefit, in this simple case of helium nuclei we were able to
test the convergence of the multiplicative cluster expansion, a crucial
treatment to be taken into account for the study of heavier systems
because the `size' of the problem is considerably reduced with respect to
the full energy computation. We note that, at least in the case of $^4$He,
the fourth-order contribution to the binding energy is non-negligible for
the (SD, $N_\beta=2$) approach.

\hfill

\hfill

This work was partially supported by DGICyT (Spain) grant PB92-0820.
M.P.~acknowledges the Instituto de Cooperaci\'on Iberoamericana (ICI,
Spain) for a fellowship.


\begin{table}
\begin{tabular}{ccccccc}
Potential & Correlation & Energy & $\alpha $ & $\beta_m$ & $a_{m,c}$ & 
$a_{m,\sigma}$ \\ \hline
S3, MS3 & SI, $N_\beta =1$ & 
$-24.4042$ & $0.8293$ & $2.0502$ & $-0.7191$ & -- \\ 
S3 & SD, $N_\beta =1$ & 
$-25.3598$ & $0.8380$ & $1.9433$ & $-0.7012$ & $0.1128$ \\ 
MS3 & SD, $N_\beta =1$ & 
$-25.3119$ & $0.8373$ & $1.9486$ & $-0.7023$ & $0.1073$ \\ \hline
S3, MS3 & SI, $N_\beta =2$ & 
$-27.2136$ & $0.5795$ & 1) $1.6206$ & $-1.5596$ & -- \\ 
&  &  &  & 2) $0.2402$ & $1.0872$ & -- \\ 
S3 & SD, $N_\beta =2$ & 
$-29.9378$ & $0.6617$ & 1) $1.4646$ & $-1.4704$ & $-0.2805$ \\ 
&  &  &  & 2) $0.3892$ & $0.9511$ & $0.4877$ \\ 
MS3 & SD, $N_\beta =2$ & 
$-29.7034$ & $0.6570$ & 1) $1.4703$ & $-1.4617$ & $-0.2454$ \\ 
&  &  &  & 2) $0.3765$ & $0.9413$ & $0.4409$ 
\end{tabular}
\caption{\label{tabS3} 
For the Afnan--Tang potentials, g.s.\ energy of a $^4$He nucleus and
optimum correlation parameters for different situations (energies are
given in MeV, $\alpha$ in fm$^{-1}$ and $\beta_m$ in fm$^{-2}$). }
\end{table}


\begin{table} 
\begin{tabular}{ccccccc}
Potential & Correlation & Energy & $\alpha $ & $\beta_m$ & $a_{m,c}$ & 
$a_{m,\sigma}$ \\ \hline 
MT~I/III, MT~V & SI, $N_\beta =1$ & 
$-29.0604$ & $0.8174$ & $3.8314$ & $-0.7934$ & -- \\ 
MT~I/III & SD, $N_\beta =1$ & 
$-29.3460$ & $0.8201$ & $3.7559$ & $-0.7853$ & $-0.0642$ \\ \hline
MT~I/III, MT~V & SI, $N_\beta =2$ & 
$-30.8752$ & $0.5089$ & 1) $3.4460$ & $-1.9090$ & -- \\ 
&  &  &  & 2) $0.1455$ & $1.4125$ & -- \\ 
MT~I/III & SD, $N_\beta =2$ & 
$-32.0107$ & $0.6077$ & 1) $3.3441$ & $-1.4139$ & $0.2230$ \\ 
&  &  &  & 2) $0.2018$ & $0.8203$ & $-0.3998$ 
\end{tabular}
\caption{\label{tabMT} 
Same considerations of table~I, for the potentials proposed by Malfliet
and Tjon. For the MT~V interaction, being of Wigner type, the SD approach
gives the same binding energy as the SI one, i.e.\ the optimum
$a_{m,\sigma}$ are zero.}
\end{table}


\begin{table}
\begin{tabular}{cccccc}
Potential & Correlation & $E^{(1)}$ & $\delta E^{(2)}$ & 
$\delta E^{(3)}$ & $\delta E^{(4)}$ \\ \hline
S3, MS3 & SI & 85.5600 & $-15.5220$ & $1.1913$ & $-0.2076$ \\ 
S3      & SD & 87.3595 & $-16.0471$ & $1.4832$ & $-0.5300$ \\ 
MS3     & SD & 87.2105 & $-16.0198$ & $1.4785$ & $-0.5151$ \\ \hline
MT~I/III, MT~V & SI & 83.1312 & $-15.6007$ & $0.5628$ & $-0.0559$ \\ 
MT~I/III       & SD & 83.6763 & $-15.7429$ & $0.6180$ & $-0.1178$ 
\end{tabular}
\caption{\label{tabdelt1} 
Contributions (given in MeV), up to all cluster orders, to the minimum
g.s.\ energy for both Afnan--Tang and Malfliet--Tjon interactions and the
optimal one-beta parametrizations. The CM corrections have not been
included. }
\end{table}


\begin{table}
\begin{tabular}{cccccc}
Potential & Correlation & $E^{(1)}$ & $\delta E^{(2)}$ & 
$\delta E^{(3)}$ & $\delta E^{(4)}$ \\ \hline
S3, MS3 & SI & 41.7835 & $-8.1422$ & $-2.4407$ & $0.0645$ \\ 
S3      & SD & 54.4792 & $-11.7120$ & $0.7360$ & $-3.4694$ \\ 
MS3     & SD & 53.7015 & $-11.5075$ & $0.5723$ & $-3.2238$ \\ \hline
MT~I/III, MT~V & SI & 32.2192 & $-6.7370$ & $-3.6119$ & $-0.1697$ \\ 
MT~I/III       & SD & 45.9464 & $-10.3470$ & $-0.0580$ & $-4.1564$ 
\end{tabular}
\caption{\label{tabdelt2} 
Same considerations of table~III, for the optimal two-beta
parametrizations. }
\end{table}

\end{document}